\documentclass[twocolumn,letter]{IEEEtran}
\usepackage[usenames,dvipsnames]{color}
\usepackage{subfigure}
\usepackage{graphicx}
\usepackage{amsmath}
\usepackage{color}
\usepackage{multicol}
\usepackage{amssymb}
\usepackage{balance}

\usepackage{amsfonts}%
\usepackage{verbatim}%
\usepackage{enumerate}%
\usepackage{psfrag}
\usepackage{epsfig}
\usepackage{multicol}
\usepackage{setspace}
\usepackage{dsfont}

\usepackage{algorithm}
\usepackage{algorithmic}
\usepackage{cite}
\usepackage{float}
\usepackage{cite}
\usepackage{fancyhdr}
   
\begin{document}
\title{Cognitive Access Protocol for Alleviating Sensing Errors in Cognitive Multiple-Access Systems}
\author{ Ahmed El Shafie \thanks {Manuscript received January 29, 2014. The associate editor coordinating the review of this letter and approving it for publication was Z. Ding.}

\thanks{A. El  Shafie is with the Wireless Intelligent Networks Center (WINC), Nile University, Smart Village, Giza, Egypt  (e-mail:ahmed.salahelshafie@gmail.com).}}
\date{}
\maketitle

%\blfootnote{Ahmed El Shafie is with the Wireless Intelligent Networks Center
%(WINC), Nile University, Smart Village, Giza, Egypt (e-mail:ahmed.salahelshafie@gmail.com).}
 \newtheorem{proposition}{Proposition}
\begin{abstract}
This letter studies a time-slotted multiple-access system with
a primary user (PU) and a secondary user (SU) sharing the same
channel resource. We propose a novel secondary access protocol which alleviates sensing errors and detects the availability of primary channels with the highest ability of detection. Under the proposed protocol, the SU may access the channel at one of a predefined instants within the time slot each of which associated with a certain access probability that changes based on the sensing outcome. There is also a possibility of accessing the channel at the beginning of the time slot without channel sensing. The optimization problem is stated such that the secondary throughput is maximized under stability of the primary queue and a constraint on the primary queueing delay. Numerical results demonstrate the beneficial gains of the proposed protocol in terms of secondary throughput.
\end{abstract}
%% \vspace{-0.1cm}
\begin{IEEEkeywords}
%%% \vspace{-0.3 cm}
Cognitive radio, secondary throughput, queue.
\end{IEEEkeywords}
%% \vspace{-0.2cm}
\section{Introduction}
%% \vspace{-0.2cm}
Cognitive radio paradigm is a promising technology to exploit the scarcity of the primary licensed spectrum. The secondary occupancy of the spectrum is an efficient solution for enhancing the spectral efficiency of the licensed spectrum. The secondary user (SU) can adapt its transmission parameters and use the licensed primary spectrum under certain quality of service requirements for the primary user (PU).

Many papers have proposed cognitive access protocols to maximize the secondary throughput under certain quality of service for the PU \cite{doha,wimob,ourletter}.
In \cite{doha}, the authors proposed a secondary access scheme on the basis of the sensing outcome. If the PU is sensed to be inactive, the SU accesses the channel with probability $1$. If the PU is sensed to be active, the SU accesses the channel with a certain access probability which is a function of
its queue length and whether it has a new packet arrival. Both PU and SU transmit with a fixed transmission
rate by employing a truncated channel inversion power control
scheme.

In \cite{wimob}, the authors investigated a simple configuration comprising of a PU and an energy harvesting SU under multipacket reception channel model. The SU probabilistically accesses and senses the primary channel. The SU may sense the channel for $\tau$ seconds or access without employing any channel sensing. If the SU decides to sense the channel, based on the sensing outcome, it changes its access probabilities. That is, the SU accesses the channel with two different probabilities based on the activity of the PU, i.e., active or inactive. Moreover, the SU can leverage the primary feedback messages to ascertain the state of the PU at the following time slot. The channel access and sensing probabilities are obtained such that the secondary throughput is maximized under certain quality of service requirement for the PU. In \cite{ourletter}, under the same configuration and channel model as in \cite{wimob}, El Shafie {\it et al.} characterized the maximum throughput of an energy harvesting SU accesses the channel randomly at the beginning of the time slot without employing any channel sensing and with a possibility of utilizing the primary feedback signal. The authors proposed simple approaches for primary parameters estimation and addressed the impact of parameters estimation errors on the secondary operation and secondary throughput.

In this letter, we consider a time-slotted multiple-access system with
a PU and an SU sharing the same
channel resource. Unlike \cite{wimob}, where the SU accesses the channel either at the beginning of the time slot or after channel sensing for $\tau$ seconds, and \cite{ourletter}, where the SU accesses the channel at the beginning of the time slot; we propose a new secondary access scheme where the SU may access the channel at one of a predefined instants within the time slot. Under the proposed protocol, the SU may access the channel at the beginning of the time slot or remain silent and gather the primary samples for $\tau$ seconds. At the instants $\tau,2\tau,\dots$, the SU decides whether to access the channel or to remain silent for $\tau$ seconds till the next decision instant. The access probability at each instant changes on the basis of the sensing outcome which is a function of the number of primary samples gathered to that instant. If the SU decides at any of the predefined instants to access the channel, its transmission continues till the far end of the time slot. Under the proposed protocol, the possibility of detecting an empty time slot is high as the SU takes a decision on channel accessing at several instants within the same time slot.

The contributions of this paper can be summarized as follows. A new access scheme for a secondary transmitter sharing the channel with a primary transmitter-receiver pair is proposed. The SU may access the channel at a specific (predefined) time instants of the time slot. The SU may access the channel at the beginning of the time slot or after channel sensing based on the declared state of the PU. The access probability when the PU is sensed to be active is, in general, different than the access probability when the PU is sensed to be inactive. The access probabilities are obtained such that the secondary throughput is maximized under stability of the primary queue and certain constraint on the primary queueing delay. We compare the proposed protocol with four recent protocols in the literature.
%% \vspace{-0.4cm}
\section{System Model}
%% \vspace{-0.2cm}
We assume a simple configuration composed of one secondary transmitter-receiver pair and one primary transmitter-receiver pair.\footnote{As
argued in the literature, e.g. \cite{doha,wimob,ourletter} and the references therein,
our system can be viewed as a subsystem within a bigger
network with multiple primary and secondary pairs using
orthogonal frequency channels.} We adopt a wireless collision channel model which means that concurrent transmissions are assumed to be lost data. The SU is assumed to be saturated and equipped with a buffer $Q_{\rm s}$ for storing its incoming traffic. The PU has a queue (buffer) $Q_{\rm p}$ for storing its incoming traffic. All queues are assumed to be of infinite capacity and contain a fixed-length packets each of $b$ bits. The channel is slotted and the length of one time slot is $T$ seconds. The
arrivals at the primary queue are independent and identically
distributed (i.i.d.) Bernoulli random variables \cite{sadek} from slot to slot with mean $\lambda_{\rm p}\in[0,1]$ packets per time slot. For similar assumptions, the reader is referred to \cite{wimob,ourletter,sadek} and the references therein.

Each receiver at the end of each time slot broadcasts a feedback signal to inform the transmitting node about the decodability status of its packet. We make use of the common assumption of error-free feedback messages, which is reasonable for short length packets as strong and low rate
codes can be employed in the feedback channel \cite{ourletter,sadek}.
When a packet is correctly received at its destination, it is then dropped from the respective transmitter's queue. In the case of packet loss due to concurrent transmission (collision) or channel outage, re-transmission of the lost data is required.

 We assume the use of an energy detector that gathers
a number of samples over a specific time duration, measures their
energy, and then compares the measured energy to a threshold
to make a decision on primary activity \cite{liang2008sensing}.

\begin{figure}
  \centering
  % Requires \usepackage{graphicx}
  \includegraphics[width=1\columnwidth]{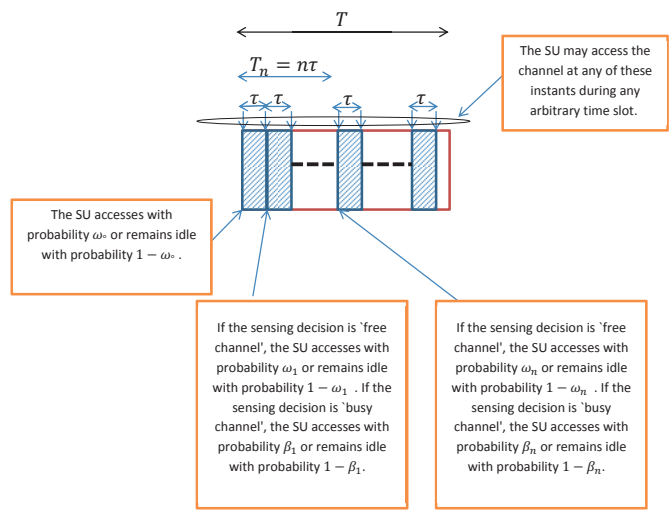}\\
   \caption{The secondary operation and time slot structure. The SU may access at $T_n=n\tau$, $n\in\{0,\tau,2\tau,\dots,\mathcal{M}\tau\}$, relative to the beginning of the slot.}\label{fig0}
   %% \vspace{-0.6cm}
\end{figure}

Let $\mathcal{M}=\lfloor T/\tau\rfloor$, where $\lfloor \mathcal{X}\rfloor$ denotes the largest integer not greater than $\mathcal{X}$. The PU accesses the channel at the beginning of the time slot if its queue is nonempty. The SU assigns $\mathcal{M}\!+\!1$ instants per time slot to be used in channel accessing (as shown in Fig. \ref{fig0}). Specifically, the SU chooses the instants $T_0, T_1, T_2,\dots,T_{\mathcal{M}}$, where $T_n=n\tau$, $n\in\{0,1,2,\dots,\mathcal{M}\}$, of the time slot to be used in channel accessing. The instant $T_0\!=\!0$ is associated with a single access probability, whereas the instant $T_\rho=\rho \tau$, $\rho\in\{1,2,\dots,\mathcal{M}\}$, is associated with two different access probabilities based on the declared activity of the PU. The choice of the access probability at $T_\rho$ depends on the sensing outcome.\footnote{The SU can select unequal $\mathcal{M}$ arbitrary instants and assign two different access probabilities to each instant. The SU can also optimize over the sensing decision duration, $\tau$. These are two possible extensions of this letter.} The operation of the SU can be summarized as follows. At the early beginning of the time slot without employing any channel sensing, the SU may access the channel with probability $\omega_\circ$ or remain idle with probability $1\!-\!\omega_\circ$. If the SU decides not to access the channel, it remains silent for $\tau$ seconds relative to the previous instant (in this case, relative to the beginning of the time slot) and starts to gather some samples from the primary signal till the next decision instant to be used for PU's activity declaration.

Based on the gathered primary samples during the interval $[0,\tau]$, the SU declares the state of the PU. If the PU is detected to be inactive, the SU accesses the channel with probability $\omega_1$; or remains idle and resumes channel sensing and primary samples gathering till instant $T_2\!=\!2\tau$ with probability $1-\omega_1$. If the PU is detected to be active, the SU accesses the channel with probability $\beta_1$; or remains idle and resumes channel sensing and primary samples gathering till instant $T_2\!=\!2\tau$ with probability $1-\beta_1$.

At $T_2\!=\!2\tau$ seconds of the time slot, based on all the gathered primary samples during $[0,2\tau]$, the SU decides on the state of the PU. If the PU is sensed to be idle, the SU accesses the channel with probability $\omega_2$. If the PU is sensed to be active, the SU accesses the channel with probability $\beta_2$. If the SU decides to remain idle, it resumes channel sensing till $3\tau$ relative to the beginning of the time slot. Generally, the SU gathers primary samples over duration $[0,\rho\tau]$, $\rho\in\{1,2,\dots,\mathcal{M}\}$, and at $T_\rho\!=\!\rho\tau$ seconds of the time slot, it accesses the channel with probability $0\le\omega_\rho\le1$ if the channel is sensed to be free; or with probability $0\le\beta_\rho\le1$ if the channel is sensed to be busy. Or it remains idle till the next decision instant with the complement probabilities of $\omega_\rho$ and $\beta_\rho$. This process continues till the end of the time slot. If the SU decides to access at any instant, the transmission continues till the far end of the time slot and the SU ceases channel sensing and primary samples gathering. The operation of the SU and the time slot structure are shown in Fig. \ref{fig0}.

All wireless links exhibit a stationary fading with
frequency non-selective Rayleigh block fading. This means
that the fading coefficient $g_{j}$ (for ${j}$ link) remains
constant during one slot, but change independently from one
slot to another according to a circularly symmetric complex
Gaussian distribution with zero mean and variance $\sigma_{j}$. The secondary link is denoted by `${\rm s}$' (link between ${\rm s}$ and its respective destination), whereas the primary link is denoted by `${\rm p}$' (link between ${\rm p}$ and its respective destination). The thermal noise at any receiver is modeled as an additive white Gaussian
noise (AWGN) with zero mean and with power spectral density $\mathcal{N}_\circ$ Watts/Hz. The channel state information is assumed to be known at the receivers only. The PU has a bandwidth of $W$ Hz. User $\ell$ transmits with power $\mathbb{P}_\ell$ Watts/Hz, $\ell\in\{\rm p,s\}$. The outage event of a link occurs
when the instantaneous capacity of the link is lower than the transmitted spectral efficiency rate.
 The correct reception (decoding) probability of a packet transmitted over the $j$th link is characterized by the complement of channel outage probability \cite{sadek,ourletter,wimob}. That is,
\begin{equation}
\overline{P^{\left(T_j\right)}_{j}}\!=\!{\rm Pr}\big\{\log_2\big(1 \!+\! \frac{\mathbb{P}_{j}}{\mathcal{N}_\circ}|g_{j}|^2\big)\!>\!\mathcal{R}_j\big\}\!=\!\exp\big(-\mathcal{N}_\circ\frac{2^{\mathcal{R}_j}\!-\!1}{\sigma_{j}
\mathbb{P}_{j}}\big)
\label{choutage}
\end{equation}
where $\mathcal{R}_j=b/(W T_j)$ is the spectral efficiency rate of node $j$ and $T_j$ is the data transmission time. Note that transmission time of the PU is $T_{\rm p}=T$, whereas the transmission time of the SU when it spends $T_n\!=\!n\tau$ seconds in channel sensing is $T_{\rm s}=T\!-\!n\tau$, where $\tau$ is the sensing duration when $n\!=\!1$.

A fundamental performance measure of a communication network is the stability of the queues.
Stability can be defined rigorously as follows.
Denote by $Q^{\left(t\right)}$ the length of queue $Q$ at the beginning of time slot $t$. Queue $Q$ is said to be stable if \cite{sadek} $\lim_{x \rightarrow \infty  } \lim_{t \rightarrow \infty  } {\rm Pr}\{Q^{\left(t\right)}<x\}=1$. We can apply Loynes theorem to check the stability of a queue \cite{sadek}. This theorem states that if the arrival process and the service process of a queue are strictly stationary, and the average service rate is greater than the average arrival rate of the queue, then the queue is stable, otherwise it is unstable.

Let $X^t_{\rm p}$ denote the number of arrivals to the primary queue $Q_{\rm p}$ at an arbitrary time slot $t$, and $Y^t_{\rm p}$ denote the number of departures of $Q_{\rm p}$ at an arbitrary time slot $t$. Based on the late arrival model, which means that an arriving packet will be blocked of getting service during its arrival time slot even if the queue is empty, the evolution of the primary queue $Q_{\rm p}$ is given by
\begin{equation}
\begin{split}
Q_{\rm p}^{t+1}=(Q_{\rm p}^t-Y_{\rm p}^t)^++X^t_{\rm p}
\end{split}
\end{equation}
where $(V)^+$ denotes $\max\{V,0\}$ and $\max\{.,.\}$ returns the maximum among the values in the argument.
%% \vspace{-0.26cm}
\section{Users throughput, primary queueing delay and Problem Formulation}\label{secx}
%% \vspace{-0.21cm}
\subsection{Primary Throughput and Queuing Delay}
%% \vspace{-0.00cm}
 Let $P_{\rm FA}^{\left(k\tau\right)}$ denote the probability that the SU's sensor generates a false alarm given that the SU sensed the channel for $k\tau$ seconds relative to the beginning of the time slot, and $P_{\rm MD}^{\left(k\tau\right)}$ denote the probability that the SU correctly detects the primary activity given that the SU sensed the channel for $k\tau$ seconds relative to the beginning of the time slot

 The service process of the primary queue is described as follows. When the SU does not access the channel at a slot, the successful packet decoding at the primary destination is characterized by the complement of channel outage between the PU and its respective receiver (link ${\rm p}$), which occurs with probability $\overline{P^{\left(T\right)}_{\rm p}}$. Taking the expectation of the primary service process, the average service rate (throughput) of the primary queue is given by
\begin{equation}
\begin{split}
\mu_{\rm p}&=\overline{P^{\left(T\right)}_{\rm p}} (1\!-\!\omega_\circ) \prod_{k=1}^{\mathcal{M}} \Big[(P^{\left(k\tau\right)}_{\rm MD}(1\!-\!\omega_k)+ \overline{P^{\left(k\tau\right)}_{\rm MD}}(1\!-\!\beta_k)\Big]
\end{split}
\end{equation}
where $\overline{P^{\left(k\tau\right)}_{\rm MD}}(1\!-\!\beta_k)$ denotes the probability that the SU detects the primary activity correctly and decides not to access the channel, $P^{\left(k\tau\right)}_{\rm MD}(1\!-\!\omega_k)$ denotes the probability of the SU misdetects the primary activity and decides not to access the channel, and $(1\!-\!\omega_\circ)\prod_{k=1}^{\mathcal{M}} \Big[(P^{\left(k\tau\right)}_{\rm MD}(1\!-\!\omega_k)+ \overline{P^{\left(k\tau\right)}_{\rm MD}}(1\!-\!\beta_k)\Big]$ means that the SU does not access the channel at any of the predefined instants within the time slot.

The probability that the primary queue being empty is given by\footnote{This formula and the delay formula in (\ref{delayform}) are obtained by solving the Markov chain modeling the primary queue.} \cite{ourletter,wimob,sadek}
\begin{equation}
\begin{split}
&{\rm Pr}\{Q_{\rm p}=0\}=1-\frac{\lambda_{\rm p}}{\mu_{\rm p}}
\end{split}
\end{equation}

The primary queueing delay is given by
\begin{equation}
\begin{split}
&D_{\rm p}=\frac{1-\lambda_{\rm p}}{\mu_{\rm p}-\lambda_{\rm p}}
\label{delayform}
\end{split}
\end{equation}
%% \vspace{-1.1cm}
\subsection{Secondary Throughput}
%% \vspace{-0.1cm}
When the PU is inactive, a packet from $Q_{\rm s}$ is served if the SU accesses the channel at any instant and the link ${\rm s}$ is not in outage. The secondary throughput is thus given by
 \begin{equation}
\begin{split}
\mu_{\rm s}&=\Big(1\!-\!\frac{\lambda_{\rm p}}{\mu_{\rm p}}\Big)\Biggr(\omega_\circ \overline{P_{\rm s}^{\left(T\right)}}+ (1\!-\!\omega_\circ)  \sum_{k=1}^{\mathcal{M}}\Biggr[ \Big(\overline{P_{\rm FA}^{\left(k\tau\right)}}\omega_k\!+\! P_{\rm FA}^{\left(k\tau\right)}\beta_k\Big) \\ & \,\,\,\,\,\,\,\,\,\ \times \prod_{\rho=1}^{k-1} \Big(\overline{P_{\rm FA}^{\left(\rho\tau\right)}} (1\!-\!\omega_\rho)\!+\! P_{\rm FA}^{\left(\rho\tau\right)} (1-\beta_\rho)\Big) \ \overline{P_{\rm s}^{\left(T\!-\!k\tau\right)}}\Biggr] \Biggr)
\end{split}
\end{equation}
where $\omega_\circ \overline{P_{\rm s}^{\left(T\right)}}$ represents the probability that the SU accesses at the beginning of the time slot and the link ${\rm s}$ is not in outage, $  \sum_{k=1}^{\mathcal{M}} \Big(\overline{P_{\rm FA}^{\left(k\tau\right)}}\omega_k\!+\! P_{\rm FA}^{\left(k\tau\right)}\beta_k\Big)$ represents the probability that the SU accesses at instant $T_k\!=\!k\tau$, $ (1\!-\!\omega_\circ) \prod_{\rho=1}^{k-1} \Big(\overline{P_{\rm FA}^{\left(\rho\tau\right)}} (1-\omega_\rho)\!+\! P_{\rm FA}^{\left(\rho\tau\right)} (1-\beta_\rho)\Big)$ represents the probability that the SU does not access at the instants preceding $k\tau$, and $\overline{P_{\rm s}^{\left(T\!-\!k\tau\right)}}$ represents the probability that the link ${\rm s}$ is not in outage given that the SU spent $k\tau$ seconds in channel sensing.

It should be noted here that as the sensing time increases, the time available for secondary data transmission decreases. Hence, the secondary channel outage increases as well. This can be seen from outage probability formula in (\ref{choutage}). It is also noted that as the sensing time increases, the reliability of the sensing outcome increases as well, i.e., the false alarm and misdetection probabilities decrease. This is actually the essence of the sensing-throughput tradeoff in cognitive radios \cite{liang2008sensing}.
    \begin{table*}
\caption{The values of false-alarm and misdetection probabilities corresponding to $[\tau,2\tau,\dots,\mathcal{M}\tau]$}
    \centering
\begin{tabular}{|@{}c@{}|c|c|c|c|c|c|c|c|c|c|}
    \hline\hline $k$& $1$&$2$&$3$&$4$&$5$&$6$&$7$&$8$&$9$&${10}$
    \\[5pt]\hline
     $P^{\left(k\tau\right)}_{\rm MD}\!=\!P^{\left(k\tau\right)}_{\rm FA}$ &$0.2$ &$0.19$ &$0.17$ &$0.15$ &$0.13$ &$0.12$ &$0.08$ &$0.05$ &$0.01$ &$0.001$ \\[5pt]\hline
%      $P^{\left(k\tau\right)}_{\rm s}$& $ 0.9981$& $0.9979$ & $0.9976$& $0.9972$& $0.9967$& $0.9959$& $0.9946$& $0.9922$& $0.9863$& $0.9545$& $0$ \\[5pt]\hline
\end{tabular}
\label{table}
%% \vspace{-0.8cm}
\end{table*}
%% \vspace{-0.3cm}
\subsection{Problem Formulation}
%% \vspace{-0.2cm}
The maximum secondary throughout is obtained via solving a constrained optimization problem. The optimization problem is stated such that the primary queue is stable and the primary queueing delay is kept lower than a specific value $\mathcal{D}$. Note that the value of $\mathcal{D}$ is application-dependent and is related to the required quality of service for the PU. The optimization problem is given by
    \begin{equation}
\begin{split}
     &  \underset{\omega_\circ,\omega_\rho,\beta_\rho \forall \rho\!\in\{\!1,2,\dots,\mathcal{M}\}}{\max.} \   \mu_{\rm s}, \\ \!\,  \,\, {\rm s.t.}    \ D_{\rm p}&\!\le\! \mathcal{D}, \ \lambda_{\rm p}\!<\! \mu_{\rm p}, \ 0\!\le\! \omega_\circ,\omega_\rho,\beta_\rho\!\le\!1 \ \forall \rho
     \label{optoptopt}
    \end{split}
    \end{equation}
where $\lambda_{\rm p}\!<\! \mu_{\rm p}$ is the primary queue stability constraint. The optimization problem (\ref{optoptopt}) is solved numerically at the SU.\footnote{We use the MatLab's fmincon to solve the optimization problem (\ref{optoptopt}) as in \cite{wimob}.} Note that the primary parameters can be known to the SU either by estimation using the primary feedback signals as in \cite{ourletter} or by cooperation between the PU and the SU.
   \begin{figure}
  \centering
  % Requires \usepackage{graphicx}
  \includegraphics[width=1\columnwidth]{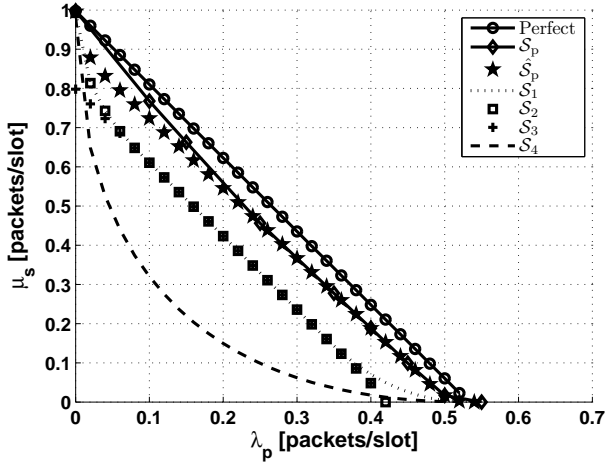}\\
   \caption{Secondary maximum throughput for each $\lambda_{\rm p}$ when $\mathcal{D}\!=\!100$ time slots. The secondary throughput under `perfect' case is given by $\mu_{\rm s}~=~(1~-~{\lambda_{\rm p}}/{\overline{P_{\rm p}^{\left(T\right)}}})\overline{P_{\rm s}^{\left(T\right)}}$ with $\lambda_{\rm p}\le\frac{\overline{P_{\rm p}^{\left(T\right)}}\mathcal{D}-1}{\mathcal{D}-1}$.}\label{fig1}
   %% \vspace{-0.4cm}
\end{figure}
%% \vspace{-0.5cm}
\section{Numerical Results and Conclusions}
%% \vspace{-0.2cm}

In this section, we present some numerical results of the proposed protocol. For comparison purposes, we present four systems: 1) the proposed protocol in \cite{wimob} where the SU accesses the channel with probability $p_f$ if the PU is sensed to be inactive, and accesses the channel with probability $p_b$ if the PU is sensed to be active, denoted by $\mathcal{S}_1$; 2) the protocol proposed in \cite{doha} where the SU accesses the channel with probability $1$ if the PU is sensed to be inactive, and with some probability $q$ if the PU is sensed to be active, denoted by $\mathcal{S}_2$; 3) the conventional access system where the SU accesses the channel with probability $1$ if and only if the PU is sensed to be inactive, denoted by $\mathcal{S}_3$; and 4) the random access without employing channel sensing discussed in \cite{ourletter} where the SU randomly accesses the channel at the beginning of the time slot, denoted by $\mathcal{S}_4$. Let $\mathcal{S}_{\rm p}$ denote the proposed protocol in this letter. We introduce a special case of $\mathcal{S}_{\rm p}$, denoted by $\mathcal{\hat{S}}_{\rm p}$, where the SU applies the proposed protocol with $\beta_{\rho}=0$ for all $\rho$. This means that the SU can access the channel randomly at the beginning of the time slot or if the PU is sensed to be inactive, otherwise, it remains idle.

   \begin{figure}
  \centering
  % Requires \usepackage{graphicx}
  \includegraphics[width=1\columnwidth]{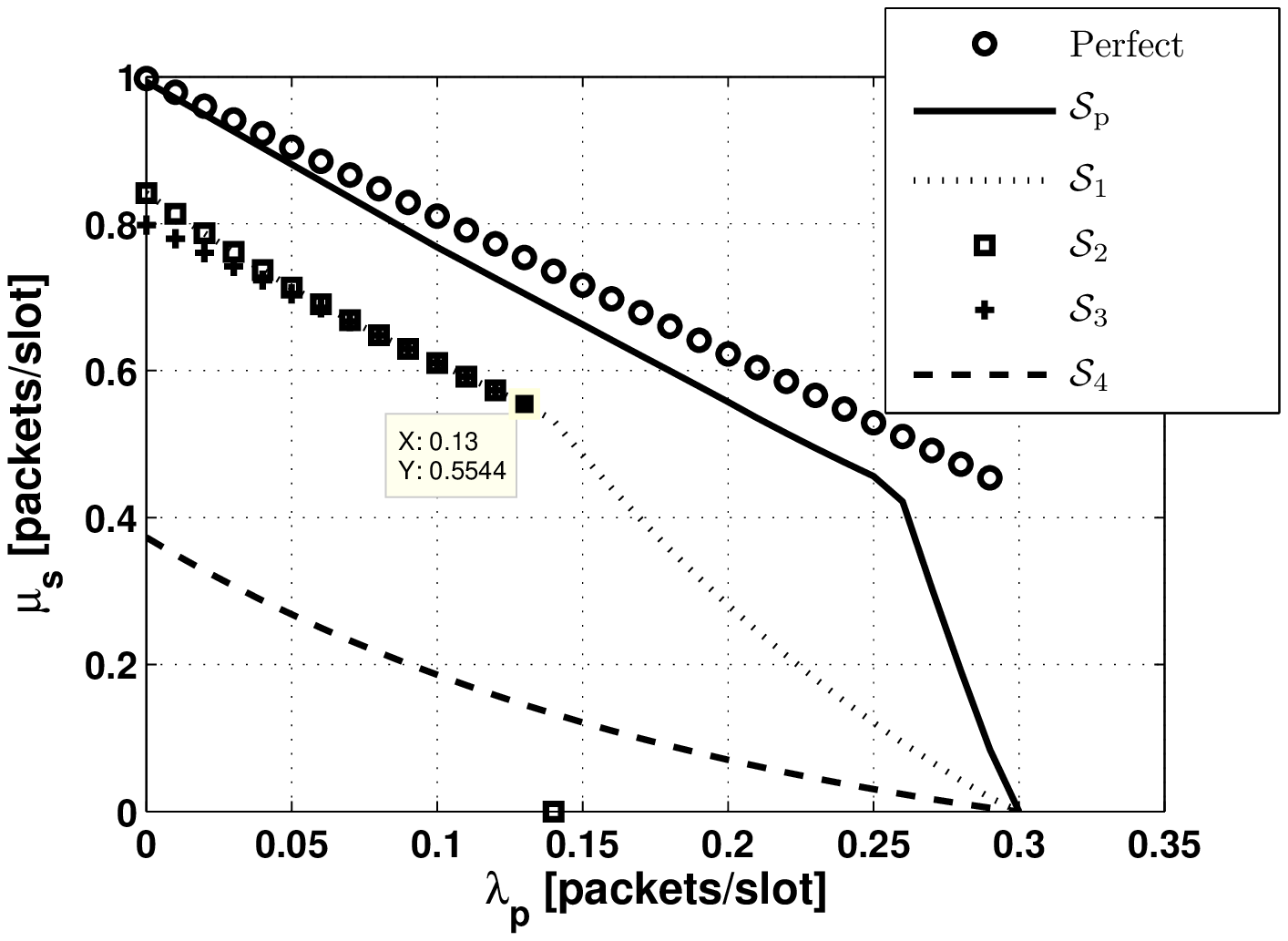}\\
   \caption{Secondary maximum throughput for each $\lambda_{\rm p}$ when $\mathcal{D}\!=\!4$ time slots. The secondary throughput under `perfect' case is given by $\mu_{\rm s}~=~\big(1~-~{\lambda_{\rm p}}/{\overline{P_{\rm p}^{\left(T\right)}}}\big)\overline{P_{\rm s}^{\left(T\right)}}$ with $\lambda_{\rm p}\le\frac{\overline{P_{\rm p}^{\left(T\right)}}\mathcal{D}-1}{\mathcal{D}-1}$.}\label{fig2}
    %% \vspace{-0.4cm}
\end{figure}
The figures are generated using $\mathcal{N}_\circ=10^{-11}$ Watts/Hz,
$\mathbb{P}_{\rm s}= 9\times 10^{-10}$ Watts/Hz,
$\mathbb{P}_{\rm p} =3\times 10^{-12}$ Watts/Hz, $W=10$ MHz, $T=0.4$ ms, $\tau=0.1 T$, $\mathcal{M}=\lfloor T/\tau\rfloor=10$, $b=1000$ bits, $\sigma_{\rm s}\!=\!\sigma_{\rm p}\!=\!1$ and Table \ref{table}. Figs. \ref{fig1} and \ref{fig2} show the maximum secondary throughout for the considered systems under two values of $\mathcal{D}$. Fig. \ref{fig1} is generated with maximum primary queueing delay $\mathcal{D}=100$ time slots, whereas Fig. \ref{fig2} is generated with maximum primary queueing delay $\mathcal{D}\!=\!4$ time slots.
We refer to the case where the SU knows perfectly without wasting any time in channel sensing that the PU is inactive in the current slot as `perfect' case. Under this case, the primary and the secondary packets correct reception are given by $\overline{P_{\rm p}^{\left(T\right)}}$ and $\overline{P_{\rm s}^{\left(T\right)}}$, respectively. This case is obviously an outer bound on what can be achieved. As shown in the figures, the secondary throughput of the proposed protocol, $\mathcal{S}_{\rm p}$, outperforms all the other protocols for all $\lambda_{\rm p}$. In addition, the figures reveal that the proposed protocol curve is close to the perfect curve. In Fig. \ref{fig1}, we note that the throughput of $\mathcal{\hat{S}}_{\rm p}$ is close to $\mathcal{S}_{\rm p}$, and they are equal at high values of $\lambda_{\rm p}$. This is because at high $\lambda_{\rm p}$ the PU will be active most of the time slots; hence, the SU does not access the channel when the PU is sensed to be active to avoid violating the PU's quality of service requirements.
 We also note that in Fig. \ref{fig2} due to the strict primary queueing delay requirement, the SU cannot access the channel for most of the primary arrival rates. That is, when $\mathcal{D}=100$ time slots, the SU can access the channel over $0\le\lambda_{\rm p}\le0.55$ packets/slot for systems $\mathcal{S}_{\rm p}$, $\mathcal{S}_{1}$ and $\mathcal{S}_{4}$ and over $0\le\lambda_{\rm p}\le0.45$ packets/slot for systems $\mathcal{S}_2$ and $\mathcal{S}_3$, whereas for $\mathcal{D}=4$ time slots, the SU cannot access the channel over $\lambda_{\rm p}\ge0.3$ packets/slot for systems $\mathcal{S}_{\rm p}$, $\mathcal{S}_{1}$ and $\mathcal{S}_{4}$ and over $\lambda_{\rm p}>0.13$ packets/slot for systems $\mathcal{S}_2$ and $\mathcal{S}_3$.

%% \vspace{-0.3cm}
\bibliographystyle{IEEEtran}
\bibliography{IEEEabrv,bibfile}
\balance

\end{document}